\documentclass[fleqn,twoside]{article}
\usepackage{espcrc2}

\usepackage{graphicx}
\usepackage[figuresright]{rotating}

   \newcommand{\HEC}{\rm HEC}
   \newcommand{\GEV}{\rm GeV}

\newcommand{\AmS}{{\protect\the\textfont2
  A\kern-.1667em\lower.5ex\hbox{M}\kern-.125emS}}

\title{The ATLAS liquid argon hadronic end-cap calorimeter: construction and 
selected beam test results}

\author{T. Barillari\address{Max-Planck-Institut f\"{u}r Physik 
        (Werner-Heisenberg-Institut) \\ 
        F\"{o}hringer Ring 6, D-80805 Muenchen, Germany}%
        \thanks{On behalf of the ATLAS \HEC\ Collaboration
          (Canada, China, Germany, Russia, Slovakia).}}
       
\begin{document}

\begin{abstract}
ATLAS has chosen for its Hadronic End-Cap Calorimeter (\HEC) the copper-liquid 
argon sampling technique with flat plate geometry and GaAs pre-amplifiers in the 
argon. The contruction of the calorimeter is now approaching completion. 
Results of production quality checks are reported and their anticipated impact 
on calorimeter performance discussed. Selected results, such as linearity, 
electron and pion energy resolution, uniformity of energy response, obtained in 
beam tests both of the Hadronic End-Cap Calorimeter by itself, and in the ATLAS 
configuration where the \HEC\ is in combination with the Electromagnetic 
End-Cap Calorimeter (EMEC) are described.

\vspace{1pc}
\end{abstract}

\maketitle

\section{INTRODUCTION}

ATLAS is a general purpose detector in construction at the ${\rm p-p}$ collider 
${\rm LHC}$ at CERN. It is designed to exploit the full discovery potential as 
given by center of mass energy of $14$ TeV and the highest luminosity of more 
than $10^{34}\,{\rm cm}^{-2}\,{\rm s}^{-1}$. The optimization of the detector 
is driven by the requirements to detect new particles and physics processes 
as e.g. the Higgs boson in the decay $H\rightarrow\gamma\gamma$, heavy 
${\rm W}-$and ${\rm Z}-$like particles or supersymmetric particles.
In the ATLAS detector the hadronic liquid argon (LAr) end-cap calorimeter 
(\HEC)~\cite{hec1,hec2} covers the pseudorapidity range $1.4<\eta<3.2$. 

To fully exploit the physics potential of ATLAS, an energy resolution 
for jets of typically 
$\sigma({\rm E})/{\rm E} = 50\%\sqrt{{\rm E\,(\GEV)}}\oplus 3\%$ 
is required for the \HEC. 
The desired linearity of the energy response measured has to stay within 
$2\%$ \cite{hec2}. 

The total energy containment up to the highest energies as 
well as an acceptable low background in the muon chambers require a calorimeter
thickness of at least $10$ interaction lengths ($\lambda$) including the 
electromagnetic end-cap calorimeter (EMEC) in front of the \HEC. 
Fast electronics is needed in order to keep the pile-up low. 

Liquid argon technology was chosen as the active medium for its robustness 
against the high radiation levels present in this forward region. 

In the following the status of the construction of the \HEC\ is discussed and
the most recent results from the $2002$ beam test measurements are presented.

\section{THE ATLAS HADRONIC END-CAP CALORIMETER}

The LAr \HEC~\cite{hec1,hec2} is a sampling calorimeter with flat copper 
absorber plates. It shares the two end-cap cryostats together with the 
EMEC and forward calorimeter (FCAL). 

The \HEC\ is structured in two wheels, the front \HEC$1$ and the rear 
\HEC$2$ wheel, placed in the cryostat behind the EMEC wheel. 
Each wheel has an outer diameter of about $4\,{\rm m}$. The length 
of \HEC$1$ (\HEC$2$) is $0.82\,{\rm m}$ ($0.96\,{\rm m}$). 
The thickness of the copper absorber plates is $25\,{\rm mm}$ for \HEC$1$ 
and $50\,{\rm mm}$ for \HEC$2$, with the first plate being half of this 
normal thickness in either case. Each wheel is made out of $32$ 
identical modules. The weight of \HEC$1$ is $67\,{\rm t}$ and that of 
\HEC$2$ is $90\,{\rm t}$.
In total $24$ gaps for \HEC$1$ and $16$ gaps for \HEC$2$ are instrumented with 
a read-out structure. Longitudinally they are read out as segments of $8$
 and $16$ gaps for \HEC$1$ and $8$ and $8$ gaps for \HEC$2$. The read-out 
structure is based on the principle of an electrostatic 
transformer ({\rm EST})~\cite{est}. The size of the liquid argon gaps between
the electrode structure is $1.85\,{\rm mm}$.

The transverse granularity, driven by the aim to reconstruct the decay 
${\rm W \rightarrow jet + jet}$ at high ${\rm p_{\perp}}$, is  
$\Delta\eta\times\Delta\phi = 0.1\times 0.1$ for the region 
$|\eta| < 2.5$ and $\Delta\eta\times\Delta\phi = 0.2\times 0.2$ beyond 
$|\eta| = 2.5$. 

The \HEC\ employs the concept of ``active pads'': the signals from individual
pads are fed into separate preamplifiers (based on highly integrated GaAs
electronics) and summed actively. The electronics boards with these IC's
are positioned directly at the module periphery, the chips are operated in
the liquid argon. Thus the input capacitances are minimized and a short
signal rise time ensured. The use of cryogenic GaAs preamplifiers and summing
amplifiers provides the optimum signal-to-noise ratio for the \HEC. The
``active pad'' electronics has been employed in more than $30$ cold tests
with and without beam. It has proven to be very reliable.

\vspace*{-0.05cm}
\section{CONSTRUCTION STATUS}

The construction of the \HEC\ is almost finished: all the modules are assembled
and have passed the cold tests, all four wheels are assembled and one full 
end-cap with \HEC$1$ and \HEC$2$ wheels is fully integrated into the end-cap 
cryostat together with the EMEC and the FCAL. The level of high voltage 
(HV) problems after integration is $\sim 0.14\%$, not causing any acceptance 
losses, but rather a few specific corrections for some channels. Only one 
out of $3072$ read-out channels is not operational, corresponding to a 
failure rate of $\sim 0.03\%$.

\section{GENERAL 2002 BEAM TEST SETUP}

The $2002$ beam test has been carried out in the H6 beam-line at the CERN SPS
providing hadrons, electrons or muons in the energy range 
$5\,\GEV \leq {\rm E} \leq 200\,\GEV$. 
The load in the liquid argon cryostat consisted 
of one EMEC module ($1/8$ of the full EMEC wheel), three \HEC$1$ modules 
($3/32$ of the full \HEC$1$ wheel) and two \HEC$2$ modules. 
Constrained by the cryostat dimension the depth of the \HEC$2$ modules were 
half of the ATLAS modules.
The impact angle of beam particles was $90^{\circ}$ with respect to the front 
plane, yielding a non-pointing geometry of the setup in $\eta$ (vertically)
unlike the ATLAS situation. 

In front of the first EMEC layer a presampler end-qcap module was placed inside 
the cryostat. It allowed studies of preshower corrections with optional 
additional material in front of the cryostat.

Finally, scintillators for triggering and timing and $4$ MWPCs with 
horizontal and vertical wire planes for beam position reconstruction 
were present further upstream in the beam line.

\section{SIGNAL RECONSTRUCTION AND CALIBRATION}

In the analysis of the beam test data,
the signal reconstruction follows closely the methods used in previous
stand-alone beam tests of the EMEC~\cite{tbemec} and the \HEC~\cite{tbhec}.
The output signals of the cold \HEC\ summing amplifiers and the raw signals from
the EMEC were carried to the front-end boards ({\rm FEB}) outside the cryostat.
Here the amplification of the EMEC signals and signal shaping of all signals
was performed. The signals were sampled at a rate of $40\,{\rm MHz}$ and stored
in a switched capacity array of the {\rm FEB}.
For each event $7$ ($16$) samples per EMEC (\HEC) channel were recorded together
with the MWPC response, trigger information and the TDC measured delay between 
the trigger and the $40\,{\rm MHz}$ sampling clock.
The raw {\rm ADC} samples were processed with an optimal filtering (OF)
method \cite{of} using $5$ event samples.

For the \HEC\ the detailed knowledge of each component in the electronics chain 
and the form of the input calibration pulse was used to determine the response 
function, which was then used to predict the shape for the physics signals.

The resulting predicted physics shapes together with the autocorrelation matrices
from noise runs was used for the computation of the OF weights. Unlike the final 
situation in ATLAS with its fixed delay between trigger and sampling clock the 
beam test trigger was asynchronous with respect to the sampling clock. The
OF weights were therefore calculated in steps of $0.5\,{\rm ns}$ in order to 
fill the $25\,{\rm ns}$ trigger window and parametrized by a $4{\rm th}$ 
order polynomial.
The accuracy for the amplitude reconstruction following this method achieved 
the level of $\pm 1\%$.  
The required conversion factor from ADC counts to {\rm nA} was obtained by 
a proper study of calibration pulse hight.

\vspace*{-0.1cm}
\section{ENERGY RECONSTRUCTION}

The total signal deposited in the calorimeter was estimated on event by event 
basis using a cell-based two-dimensional topological clustering algorithm. 

The electronic noise $\sigma_{n}$ (in nA) for each cell was obtained by studying 
either muon data or time samples located outside the physics pulse time region. 

Each cluster consisted of at least one cell with a signal-to-noise ratio
above 4 ($E > 4\sigma_{n}$). A threshold on the absolute value of the
signal-to-noise ratio, $|E| > 2\sigma_{n}$, was applied to all other
cells. They were included in the cluster if they shared at least one
edge with a cluster member cell satisfying $|E| > 3\sigma_{n}$.  The
symmetric cuts on the cell and neighbor level avoided biases due to
electronics noise. Two super-clusters for the EMEC and the \HEC\ were
defined by summing all cluster signals in the EMEC and the \HEC\,
respectively. For the \HEC\ the signals in the $3^{\rm rd}$ layer were
multiplied by $2$ in order to account for the $50\,\%$ smaller sampling
fraction.

\section{RESPONSE TO ELECTRONS}

From Monte Carlo simulations of the beam test configuration the
leakage outside the EMEC was found to be very small for electrons and
has been neglected.  Therefore the ratio of the known beam energy
($6-150\,{\rm GeV}$) and the sum of all signals in the EMEC in nA
defined the electromagnetic scale factor, $\alpha_{\rm em}^{\rm EMEC}
= 0.430\pm0.001\,{\rm MeV}/{\rm nA}$, where the error is here statistical
only. The variation with energy tests the linearity in the energy
range considered and was found to be better than $0.5\,\%$.

The energy resolution for electrons has been studied using the super-cluster
mentioned above.

In data the energy resolution was found to be $\sigma(E)/E =
(0.121\pm0.002)/\sqrt{E/{\rm GeV}} \oplus 0.004\pm0.001$ after noise
subtraction.  The noise $\sigma_{\rm n} \simeq 0.2-0.3\,{\rm GeV}$
varies with energy due to the non-fixed cluster size. 
GEANT3~\cite{Geant3,Geant3HEC} and GEANT4~\cite{Geant4}) based Monte Carlo 
simulations are in good agreement with the data and yield for GEANT3 an 
energy resolution
$\sigma(E)/E = (0.093\pm0.006)/\sqrt{E/{\rm GeV}} \oplus 0.008\pm0.001$
while GEANT4 yield a 
$\sigma(E)/E =(0.106\pm0.007)/\sqrt{E/{\rm GeV}} \oplus 0.007\pm0.002$.
Further tuning is required for ATLAS to improve the fair agreement with the data.

\section{RESPONSE TO PIONS}

The electromagnetic scale for the \HEC\ was taken from the previous
stand-alone beam test~\cite{HECNIM}, $\alpha_{\rm em}^{\rm HEC} =
3.27\pm0.03\,{\rm MeV}/{\rm nA}$, taking the modified electronics into
account.  A good agreement of the total visible energy in the EMEC and
\HEC\ for pions with Monte Carlo simulations based either on GEANT3 or
the quark-string-gluon-plasma (QGSP) model of GEANT4 is observed, while
the GEANT4 low-and-high-energy-pion-parameterization (LHEP) model
deviate largely from data.

\section{WEIGHTING}

The non-compensating nature of the two calorimeters makes weighting of
hadronic energy deposits necessary. A cell based weighting method
which was successfully used in previous
experiments~\cite{H1det,H1larg} needs a detailed simulation on
the cell level, which is not yet available for ATLAS. Therefore a
more coarse weighting scheme on the super-cluster level has been
applied.

With the leakage outside the detector volume as predicted by the Monte
Carlo and the known beam energy 6 weights (3 for the EMEC and 3 for
the \HEC) have been fitted from the two super-cluster energies and their
energy density, leading to the weighted energies $E_{\rm w} = E_{\rm
em}\left(C_1\cdot\exp\left[-C_2 \cdot E_{\rm em}/V\right]+C_3\right)$.

\begin{figure}[htbp]\vspace*{-0.7cm}
 \centering
  \resizebox{0.35\textwidth}{!}{%
  \includegraphics{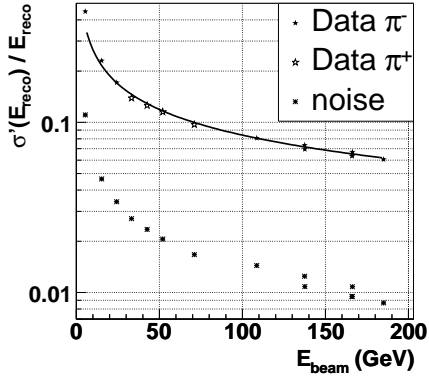}}
    \vspace*{-1.cm}
    \caption{Energy dependence of the energy resolution for pion data using
    the cluster weighting approach. The subtracted noise is shown explicitly.
    The line shows the result of the fit to the data.}
    \label{fig:erpion}
\vspace*{-0.6cm}
\end{figure}

The noise subtracted energy resolution for pions, 
$\sigma(E)/E = (0.841\pm0.003)/\sqrt{E/{\rm GeV}} \oplus 0\pm0.003$,
see Fig.~\ref{fig:erpion}, 
is slightly worse than expected from Monte Carlo.
The GEANT3 simulation predicts a
$\sigma(E)/E = (0.733\pm0.005)/\sqrt{E/{\rm GeV}} \oplus 0\pm0.0$
and a GEANT4 LHEP and QGSP simulations predict an energy resolution
sampling term of $(74.0\pm0.5)\%\,\sqrt{\rm GeV}$ and 
$(72.3\pm0.9)\%)\sqrt{\rm GeV}$,
respectively. In general the GEANT4 models seem closer to the data, 
but neither QGSP and LHEP give an optimal description. 
The different energy dependence of the GEANT4 predictions is also reflected 
in non-vanishing constant terms: $(4.1\pm0.1)\%$ for LHEP and $(2.5\pm0.3)\%$ 
for QGSP.

\begin{figure}[htbp]\vspace*{-0.5cm}
 \centering
  \resizebox{0.37\textwidth}{!}{%
  \includegraphics{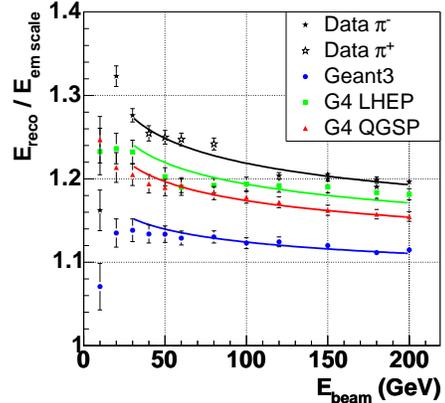}}
    \vspace*{-1.1cm}
    \caption{$e/\pi$-ratio as obtained from the cluster weighting function.
    Shown is the energy dependence for the data as well as for the different
    MC models. The lines show the results of fits to the energy dependence.}
    \label{fig:eoverpi}
\vspace*{-0.5cm}
\end{figure}

The ratio of the combined weighted energy of EMEC and \HEC\ over the
combined electromagnetic energy yield the effective ${\rm
e}/\pi$-ratio for the end-cap calorimeters ranging from $1.32$ at
$20\,{\rm GeV}$ to $1.19$ at $200\,{\rm GeV}$ for pions.  
Fig.~\ref{fig:eoverpi}
shows this ratio for the data and the different MC models. The energy 
dependence is in all cases rather similar. However the MC predictions
are here substantially below the data, with GEANT3 being especially low.

\end{document}